\documentclass[submission]{eptcs}
\providecommand{\event}{SCSS 2021} 
\usepackage{breakurl}             
\usepackage{underscore}           
\usepackage{amsthm,amsmath}
\usepackage{xcolor}

\usepackage{fancyvrb}
\usepackage[all]{xy}
\SelectTips{cm}{}
\usepackage{amssymb}

\newtheorem{theorem}{Theorem}[section]
\newtheorem{proposition}[theorem]{Proposition}
\theoremstyle{definition}
\newtheorem{definition}[theorem]{Definition}
\newtheorem{remark}[theorem]{Remark}
\newtheorem{example}[theorem]{Example}

\newcommand{\sparql}{\mathrm{SPARQL}} 
\newcommand{\sql}{\mathrm{SQL}} 
\newcommand{\gral}{\mathrm{GrAL}} 

\newcommand{\Ll}{{\mathcal{L}}} 
\newcommand{\mC}{{\mathcal{C}}} 
\newcommand{\V}{{\mathcal{V}}} 
\newcommand{\T}{{\mathcal{T}}} 

\newcommand{\hsp}{\null\hspace{1pc}}
\newcommand{\To}{\Rightarrow} 
\newcommand{\lwedge}{\,\wedge\,} 

\newcommand{\Ulm}{\mathit{Match}} 
\newcommand{\ulm}{{\underline{m}}}  
\newcommand{\ulp}{{\underline{p}}}
\newcommand{\ulc}{{\underline{c}}}
\newcommand{\ulf}{{\underline{f}}}  
\newcommand{\ulq}{{\underline{q}}} 

\newcommand{\tbl}{\mathit{Tab}}
 
\newcommand{\expr}{e} 
\newcommand{\op}{\mathit{op}} 
\newcommand{\Op}{\mathit{Op}} 
\newcommand{\agg}{\mathit{agg}} 
\newcommand{\Agg}{\mathit{Agg}} 
\newcommand{\gp}{\mathit{gp}} 
\newcommand{\true}{\mathit{true}} 
\newcommand{\false}{\mathit{false}} 
\newcommand{\ev}{\mathit{ev}}
\newcommand{\elem}{\mathit{elem}}
\newcommand{\err}{\bot} 
\newcommand{\deno}[1]{[[#1]]} 
\newcommand{\bag}[1]{\{\!|#1|\!\}} 
\newcommand{\var}[2]{\mathit{var}(#1,#2)} 

\newcommand{\se}[1]{{[#1]}} 
\newcommand{\sem}[2]{{[[#1]]_{#2}}} 
\newcommand{\seg}[2]{{{#2}^{(#1)}}} 

\newcommand{\opn}[1]{\mbox{{\it #1}}} 
\newcommand{\clause}[1]{\mbox{{\rm #1}}} 
\newcommand{\Card}[1]{\mathit{Card}(#1\,)}  
\newcommand{\Result}{\mathit{Result}}  
\newcommand{\param}{\mathit{param}}

\newcommand{\hide}[1]{} 

\title{Querying RDF Databases with Sub-CONSTRUCTs}
\author{Dominique Duval
\institute{LJK - Univ. Grenoble Alpes}
\email{dominique.duval@univ-grenoble-alpes.fr}
\and
Rachid Echahed
\institute{LIG - Univ. Grenoble Alpes}
\email{rachid.echahed@imag.fr}
\and Fr\'ed\'eric Prost
\institute{LIG - Univ. Grenoble Alpes}
\email{frederic.prost@univ-grenoble-alpes.fr}
}

\def\titlerunning{Querying RDF Databases with Sub-CONSTRUCTs}
\def\authorrunning{D. Duval, R. Echahed \& F. Prost}

\begin{document}
\maketitle

\begin{abstract}

Graph query languages feature mainly two kinds of queries when applied
to  a graph  database: those  inspired by  relational databases  which
return tables  such as  SELECT queries and  those which  return graphs
such as CONSTRUCT queries in SPARQL. The latter are object of study in
the present paper. For this purpose,  a core graph query language GrAL
is defined with focus on CONSTRUCT  queries.  Queries in GrAL form the
final   step  of   a  recursive   process  involving   so-called  GrAL
patterns. By evaluating  a query over a graph one  gets a graph, while
by evaluating a pattern  over a graph one gets a  set of matches which
involves both  a graph and  a table. CONSTRUCT queries are  based on
CONSTRUCT patterns, and sub-CONSTRUCT patterns  come for free from the
recursive definition of  patterns.  The semantics of GrAL  is based on
RDF  graphs with  a slight  modification which  consists in  accepting
isolated nodes.  Such an extension  of RDF graphs eases the definition
of  the evaluation  semantics, which  is mainly  captured by  a unique
operation called  Merge.  Besides, we  define aggregations as  part of
GrAL  expressions, which  leads  to an  original  local processing  of
aggregations.
\end{abstract}
\maketitle

\section{Introduction}
Graph database query languages are becoming ubiquitous. In contrast to
classical relational databases where SQL language is a standard,
different languages \cite{AnglesABHRV17} have been proposed for
querying graph databases, like SPARQL \cite{sparql} or Cypher
\cite{cypher}.
Among the most popular models for representing graph databases, one
may quote for instance the \emph{sets of triples} (or \emph{RDF
graphs} \cite{rdf}) used by $\sparql$ or the
\emph{property graphs} used by Cypher.
In addition to the lack of a standard model to represent graph
databases, there are different kinds of queries in the context of
graph query languages. One may essentially distinguish two classes of
queries: those inspired by relational databases which return tables
such as SELECT queries and those which return graphs such as CONSTRUCT
queries in SPARQL. Such CONSTRUCT queries are graph-to-graph queries
specific to graph databases.

The graph-to-graph queries received less attention than the
graph-to-table queries.  For instance, for the SPARQL language, a
semantics of SELECT queries and subqueries is proposed in \cite{KKC},
a semantics of CONSTRUCT queries in \cite{Kostylevetal2015} and a
semantics of CONSTRUCT queries with nested CONSTRUCT queries in FROM
clauses in \cite{AnglesG11,PolleresRK16}, where the outcome of a
subCONSTRUCT is a graph. All these works consider graphs as sets of
triples.  Unfortunately, such a definition of graphs prevents having a
uniform semantics of all patterns and in particular for BIND patterns.

In this paper, we focus on graph-to-graph queries and subqueries for
RDF graphs and we propose a new semantics for CONSTRUCT subqueries which
departs from the one in \cite{AnglesG11,PolleresRK16}. First, we
propose to change slightly the definition of graphs by allowing
isolated nodes. Indeed, this new definition of graphs allows us to
have a uniform semantics of all patterns. 
In fact, we define CONSTRUCT \emph{subpatterns} rather than
CONSTRUCT subqueries.
For this purpose, we introduce a core query language $\gral$ based on
RDF graphs.
The syntactic categories of $\gral$ include both queries and patterns. 
When  evaluating a CONSTRUCT query over a graph one gets a graph,
whereas when evaluating a CONSTRUCT pattern over a graph one gets a set of
matches which involves both variable assignments and a graph.
In fact, a CONSTRUCT query first acts as a CONSTRUCT pattern and then 
returns only the constructed graph.
As the definition of patterns is recursive, CONSTRUCT subpatterns
are obtained for free. 

In order to define the semantics of $\gral$, we introduce an
algebra of operations over sets of matches,
where a match is a morphism between graphs.
We propose to base the semantics of $\gral$ upon an algebra on sets of matches,
like the semantics of $\sql$ is based upon relational algebra.
All operations in our algebra essentially derive from a
unique operation called \emph{Merge}, which generalizes the well-known
\emph{Join} operation. As stated earlier, 
we consider graphs consisting of classical RDF triples possibly augmented
with some additional isolated nodes. This slight extension helps
formulating the semantics of patterns and queries without using some
cumbersome notations to handle, for instance, environments defined by
variable bindings.  
The proposed algebra is used to define an evaluation
semantics for $\gral$. As for aggregations, they are handled
locally inside expressions. The semantics of the various
patterns and queries is uniform, as it is based on instances of the
\emph{Merge} operation.

The paper is organized as follows. Section~\ref{sec:algebra}
introduces the algebra designed to express the semantics of the query
language $\gral$.  In Section~\ref{sec:gral}, the language $\gral$ is
defined by its syntax and semantics.
Concluding remarks are given in Section~\ref{sec:conclusion}.

\section{The Graph Query Algebra}
\label{sec:algebra}

The Graph Query Algebra is a family of operations which are used in
Section~\ref{sec:gral} for defining the evaluation of queries in the 
Graph Algebraic Query Language $\gral$.
Graphs and matches are introduced in Section~\ref{ssec:algebra-graphs},
then operations on sets of matches are defined
in Section~\ref{ssec:algebra-operation}.

\subsection{Graphs and matches}
\label{ssec:algebra-graphs}
In this paper, graphs are kinds of generalised RDF graphs that may contain
isolated nodes. 
Let $\Ll$ be a set, called the set of \emph{labels}, 
union of two disjoint sets $\mC$ and $\V$, called respectively
the set of \emph{constants} and the set of \emph{variables}.

\begin{definition}[graph] 
\label{def:algebra-graph}
Every element $t=(s,p,o)$ of $\Ll^3$ is called a \emph{triple}
and its members $s$, $p$ and $o$ are called respectively 
the \emph{subject}, \emph{predicate} and \emph{object} of $t$.
A \emph{graph} $X$ is made of a subset $X_N$ of $\Ll$
called the set of \emph{nodes} of $X$ and a subset $X_T$ of $\Ll^3$ 
called the set of \emph{triples} of $X$, such that the subject and the object
of each triple of $X$ are nodes of $X$.
The nodes of $X$ which are neither a subject nor an object are called
the \emph{isolated nodes} of $X$.
The set of \emph{labels} of a graph $X$ is the subset $\Ll(X)$
of $\Ll$ made of the nodes and predicates of $X$, then 
$\mC(X)=\mC\cap\Ll(X)$ and $\V(X)=\V\cap\Ll(X)$.
Given two graphs $X_1$ and $X_2$, 
the graph $X_1$ is a \emph{subgraph} of $X_2$, written $X_1 \subseteq X_2$,
if $(X_1)_N \subseteq (X_2)_N$ and $(X_1)_T \subseteq (X_2)_T$,
then obviously $\Ll(X_1) \subseteq \Ll(X_2)$.
The \emph{union} $X_1\cup X_2$ is the graph 
defined by $(X_1\cup X_2)_N = (X_1)_N \cup (X_2)_N$ and
$(X_1\cup X_2)_T = (X_1)_T \cup (X_2)_T$,
then $\Ll(X_1\cup X_2)=\Ll(X_1)\cup \Ll(X_2)$.
\end{definition}

We will not use the \emph{intersection} $X_1\cap X_2$, which could be 
defined by $(X_1\cap X_2)_N = (X_1)_N \cap (X_2)_N$ and
$(X_1\cap X_2)_T = (X_1)_T \cap (X_2)_T$: then 
the intersection of two graphs without isolated nodes might have
isolated nodes and $\Ll(X_1\cap X_2)$ might be strictly smaller than
$\Ll(X_1)\cap \Ll(X_2)$, as for instance when $X_1=\{(x,y,z)\}$ and
$X_2=\{(y,z,x)\}$ so that $X_1\cap X_2=\{x\}$.

\begin{example}
  \label{ex:database}

  We introduce here a toy database representing a simplified view of a
  social media network. We will use it as a running example to
  illustrate various definitions.  The network consists in {\em
    authors} {\em publishing} {\em messages}.  Each message is {\em
    timestamped at} a certain {\em date} (a day).  A message can {\em
    refer to} other messages and an author may {\em like} a message.
  An instance of such a network is described by the following graph
  $G_0$ (written ``\`a la'' RDF):

\begin{center}
  $G_0=$
  {\tt
\begin{tabular}{|l|}
  \hline
  auth1 publishes mes1 . auth1 publishes mes2 . \\
  auth2 publishes mes3 . auth3 publishes mes4 . auth3 publishes mes5 . \\ 
  mes1 stampedAt date1 . mes2 stampedAt date2 . \\
  mes3 stampedAt date1 . mes4 stampedAt date4 . mes5 stampedAt date4 . \\
  mes3 refersTo mes1   . mes4 refersTo mes1   . mes4 refersTo mes2 .  \\
  auth1 likes mes3     . auth1 likes mes4     . auth1 likes mes5 . \\
  auth2 likes mes1     . auth2 likes mes4 \\
  \hline
\end{tabular}
}
\end{center}

The meaning of $G_0$ is that author {\tt auth1} has published messages
{\tt mes1} and {\tt mes2}, which have been stamped respectively at dates
{\tt date1} and {\tt date2}, etc.
\end{example}

\begin{definition}[match]
\label{def:algebra-match}
A \emph{match} $m$ from a graph $X$ to a graph $G$, denoted $m:X\to G$, 
is a function from $\Ll(X)$ to $\Ll(G)$
which \emph{preserves nodes} and \emph{preserves triples} 
and which \emph{fixes} $\mC$, in the sense that 
$m(X_N)\subseteq G_N$, $m^3(X_T)\subseteq G_T$ and
$m(x)=x$ for each $x$ in $\mC(X)$.
The set of all matches from $X$ to $G$ is denoted $\Ulm(X,G)$. 
An \emph{isomorphism} of graphs is an invertible match. 
\end{definition}

When $n$ is an isolated node of $X$ then the node $m(n)$ does not
have to be isolated in $G$.
A match $m:X\to G$ determines two functions 
$m_N:X_N\to G_N$ and $m_T:X_T\to G_T$, restrictions of $m$ and $m^3$ 
respectively. 
A match $m:X\to G$ is an isomorphism if and only if 
both functions $m_N:X_N\to G_N$ and $m_T:X_T\to G_T$ are bijections.
This means that a function $m$ from $\Ll(X)$ to $\Ll(G)$
is an isomorphism of graphs if and only if
$\mC(X)=\mC(G)$ with $m(x)=x$ for each $x\in\mC(X)$
and $m$ is a bijection from $\V(X)$ to $\V(G)$:
thus, $X$ is the same as $G$ up to variable renaming.
It follows that the symbol used for naming a variable does not matter
as long as graphs are considered only up to isomorphism.

\begin{definition}[image of a graph by a function] 
\label{def:algebra-image}
Let $X$ be a graph.
Every function $f:\V(X)\to\Ll$ can be extended in a unique way as a function
$f':\Ll(X)\to\Ll$ that fixes $\mC$. 
The \emph{image} $f(X)$ of $X$ by $f$ is the graph made of
the nodes $f'(n)$ for $n\in X_N$ and the triples $(f')^3(t)$ for $t\in X_T$.
The function $f$ can be extended in a unique way as a match
$f^{\sharp}:X\to f(X)$.
\end{definition}

\begin{definition}[compatible matches] 
\label{def:algebra-compatible}  
Two matches $m_1:X_1\to G_1$ and $m_2:X_2\to G_2$ are \emph{compatible},
written as $m_1\sim m_2$, if $m_1(x)=m_2(x)$ for each $x\in\V(X_1)\cap\V(X_2)$. 
Given two compatible matches $m_1:X_1\to G_1$ and $m_2:X_2\to G_2$, 
let $m_1\bowtie m_2 : X_1\cup X_2 \to G_1\cup G_2$
denote the unique match such that $m_1\bowtie m_2\sim m_1$
and $m_1\bowtie m_2\sim m_2$ 
(which means that $m_1\bowtie m_2$ coincides with $m_1$ on $X_1$
and with $m_2$ on $X_2$). 
\end{definition}

We will see in Section~\ref{sec:gral} that the execution of a query in $\gral$
is a graph-to-graph transformation,
which main part is a graph-to-set-of-matches transformation.

\begin{definition}[set of matches, assignment table] 
\label{def:algebra-table}
Let $X$ and $G$ be graphs. A set $\ulm$ of matches, 
all of them from $X$ to $G$, is denoted $\ulm:X\To G$.
The \emph{assignment table} $\tbl(\ulm)$ of $\ulm$ is the two-dimensional
table with 
the elements of $\V(X)$ in its first row, then one row for each $m$ in $\ulm$, 
and the entry in row $m$ and column $x$ equals to $m(x)$.
\end{definition}

Thus, the assignment table $\tbl(\ulm)$ describes the set of functions
$\ulm|_{\V(X)}:\V(X)\To\Ll$,
made of the functions $m|_{\V(X)}:\V(X)\to\Ll$ for all $m\in\ulm$.
The set of matches $\ulm:X\To G$ is determined by the graphs $X$ and $G$
and the assignment table $\tbl(\ulm)$. This property
is used hereafter to describe some examples.

\begin{example}
  \label{ex:set-of-matches}
  Here are some examples of matches in the graph $G_0$
  (defined in Example~\ref{ex:database}).
  
  \begin{itemize}
    
    \item
  Let $P_{\it ps}$ be the following graph
  (written ``\`a la'' SPARQL: variable names begin with ``{\tt ?}''):
  \begin{center}
    $P_{\it ps}=$
{\tt
\begin{tabular}{|l|}
  \hline
   ?a publishes ?m . ?m stampedAt ?d \\
  \hline
\end{tabular}
} 
\end{center}
  The set $\ulm_{\it ps}$ of all matches from $P_{\it ps}$ to $G_0$ is: 
$$ \ulm_{\it ps}:P_{\it ps} \To G_0 \;\mbox{ with }\;
  \tbl(\ulm_{\it ps}) = \begin{array}{|l|l|l|}
  \hline
  {\tt ?a} & {\tt ?m} & {\tt ?d} \\
  \hline
  {\tt auth1} & {\tt mes1} & {\tt date1} \\
  {\tt auth1} & {\tt mes2} & {\tt date2} \\
  {\tt auth2} & {\tt mes3} & {\tt date1} \\
  {\tt auth3} & {\tt mes4} & {\tt date4} \\
  {\tt auth3} & {\tt mes5} & {\tt date4} \\
  \hline
  \end{array} 
  $$
  
    \item
  Let $P_{\it pl}$ be the graph:
  \begin{center}
    $P_{\it pl}=$
{\tt
\begin{tabular}{|l|}
  \hline
   ?a1 publishes ?m . ?a2 likes ?m \\
  \hline
\end{tabular}
} 
\end{center}
The set $\ulm_{\it pl}$ of all matches from $P_{\it pl}$ to $G_0$ is: 
$$ \ulm_{\it pl}:P_{\it pl} \To G_0 \;\mbox{ with }\;
  \tbl(\ulm_{\it pl}) = \begin{array}{|l|l|l|}
  \hline
  {\tt ?a1} & {\tt ?m} & {\tt ?a2} \\
  \hline
  {\tt auth1} & {\tt mes1} & {\tt auth2} \\
  {\tt auth2} & {\tt mes3} & {\tt auth1} \\
  {\tt auth3} & {\tt mes4} & {\tt auth1} \\
  {\tt auth3} & {\tt mes4} & {\tt auth2} \\
  {\tt auth3} & {\tt mes5} & {\tt auth1} \\
  \hline
  \end{array} 
  $$
  
    \item
  Let $P'_{\it pl}$ be the following subgraph of $P_{\it pl}$,
  made of two isolated nodes:
\begin{center}
  $P'_{\it pl}=$
  {\tt
\begin{tabular}{|l|}
  \hline
   ?a1 . ?a2 \\
  \hline
\end{tabular}
} 
\end{center}
The subset $\ulm'_{\it pl}$ of $\ulm_{\it pl}$ made of the
restrictions to $P'_{\it pl}$ of the matches in $\ulm_{\it pl}$ is: 
$$ \ulm'_{\it pl} : P'_{\it pl} \To G_0  \;\mbox{ with }\;
  \tbl(\ulm'_{\it pl}) = \begin{array}{|l|l|}
  \hline
  {\tt ?a1} & {\tt ?a2} \\
  \hline
  {\tt auth1} & {\tt auth2} \\
  {\tt auth2} & {\tt auth1} \\
  {\tt auth3} & {\tt auth1} \\
  {\tt auth3} & {\tt auth2} \\
  \hline
  \end{array} 
  $$
  
    \item
  Let $P_{\it prp}$ be the graph:
  \begin{center}
    $P_{\it prp}=$
{\tt
\begin{tabular}{|l|}
  \hline
   ?a1 publishes ?m1 . ?m1 refersTo ?m2 . ?a2 publishes ?m2 \\
  \hline
\end{tabular}
} 
\end{center}
The set $\ulm_{\it prp}$ of all matches from $P_{\it prp}$ to $G_0$ is: 
$$ \ulm_{\it prp}:P_{\it prp} \To G_0 \;\mbox{ with }\;
  \tbl(\ulm_{\it prp}) = \begin{array}{|l|l|l|l|}
  \hline
  {\tt ?a1} & {\tt ?m1} & {\tt ?m2} & {\tt ?a2} \\
  \hline
  {\tt auth2} & {\tt mes3} & {\tt mes1} & {\tt auth1} \\
  {\tt auth3} & {\tt mes4} & {\tt mes1} & {\tt auth1} \\
  {\tt auth3} & {\tt mes4} & {\tt mes2} & {\tt auth1} \\
  \hline
  \end{array} 
  $$
  \end{itemize}
  
\end{example}

\begin{definition}[image of a graph by a set of functions] 
\label{def:algebra-image-s}
The \emph{image} of a graph $X$ by a set of functions
$\ulf$ from $\V(X)$ to $\Ll$, denoted $\ulf(X)$, 
is the graph union of the graphs $f(X)$ for every $f$ in $\ulf$.
Every set of functions $\ulf:\V(X)\To\Ll$ can be extended in a unique way as a
set of matches $\ulf^{\sharp}:X\To\ulf(X)$.
\end{definition}

\begin{remark}[about RDF graphs]
\label{rem:algebra-rdf}
RDF graphs \cite{rdf} are graphs (as in Definition~\ref{def:algebra-graph})
without isolated nodes, where constants are either IRIs
(Internationalized Resource Identifiers) or literals
and where all predicates are IRIs and only objects can be literals.
Blank nodes in RDF graphs are the same as variable nodes in our graphs. 
Thus an isomorphism of RDF graphs, as defined in~\cite{rdf},
is an isomorphism of graphs as in Definition~\ref{def:algebra-match}.
\end{remark}

\subsection{Operations on sets of matches}
\label{ssec:algebra-operation}
In this Section we introduce some operations on sets of matches 
which are used in Section~\ref{sec:gral} for defining the semantics
of $\gral$.
The prominent one is the \emph{merging} operation
(Definition~\ref{def:algebra-merge}), which is a kind of generalized 
\emph{joining} operation (see Definition~\ref{def:algebra-opn}).
Other basic operations are the simple \emph{restriction} and \emph{extension}
operations
(Definitions~\ref{def:algebra-restrict} and~\ref{def:algebra-extend}). 
Then, these basic operations are combined in order to get some
derived operations (Definition~\ref{def:algebra-opn}). 

\begin{definition}[Merge]
\label{def:algebra-merge}
Let $\ulm:X\To G$ be a set of matches and $\ulp_m : Y\To H_m$
a family of sets of matches indexed by $m\in\ulm$,
and let $H=\cup_{m\in\ulm}H_m$.
The \emph{merging} of $\ulm$ along the family $(\ulp_m)_{m\in\ulm}$ is 
the set of matches $m\bowtie p$ for every $m\in\ulm$ and
every $p\in\ulp_m$ compatible with~$m$:
\\ \hsp $\opn{Merge}(\ulm,(\ulp_m)_{m\in\ulm}) 
=\{ m\bowtie p \mid m\in\ulm \lwedge p\in\ulp_m \lwedge m\sim p\}
: X\cup Y \To G\cup H$.
\end{definition}

Let $\ulq=\opn{Merge}(\ulm,(\ulp_m)_{m\in\ulm})$, then $\ulq$ is made of
a match $m\bowtie p$ for each pair $(m,p)$ with
$m\in\ulm$ and $p\in\ulp_m$ compatible with $m$
(so that for each $m$ in $\ulm$ the number of $m\bowtie p$ in $\ulq$
is between~0 and $\Card{\ulp_m}$). 
The match $m\bowtie p:X\cup Y \to G\cup H$ is such that
$m\bowtie p(x)=m(x)$ when $x\in X$ and $m\bowtie p(y)=p(y)$ when $y\in Y$,
which is unambiguous because of the compatibility condition.

\begin{example}
  \label{ex:merge} 
  Here are some examples of merging, based on the sets of matches 
  in Example~\ref{ex:set-of-matches}.
  \begin{itemize}
    \item
      Let $\ulp'_{\it pl}$ be similar to $\ulm'_{\it pl}$, up to renaming
      the variables {\tt ?a1} and {\tt ?a2} as {\tt ?a2} and {\tt ?a1},
      respectively, so that $\ulp'_{\it pl}$ is:
      $$ \ulp'_{\it pl} : P'_{\it pl} \To G_0  \;\mbox{ with }\;
  \tbl(\ulp'_{\it pl}) = \begin{array}{|l|l|}
  \hline
  {\tt ?a1} & {\tt ?a2} \\
  \hline
  {\tt auth2} & {\tt auth1} \\
  {\tt auth1} & {\tt auth2} \\
  {\tt auth1} & {\tt auth3} \\
  {\tt auth2} & {\tt auth3} \\
  \hline
  \end{array} 
$$
  For each match $m$ in $\ulm'_{\it pl}$ let $\ulp_m=\ulp'_{\it pl}$,
  which does not depend on $m$.
  Then $\opn{Merge}(\ulm'_{\it pl},(\ulp_m)_{m\in\ulm'_{\it pl}})$
  is denoted simply $\opn{Join}(\ulm'_{\it pl},\ulp'_{\it pl})$
  (as in Definition~\ref{def:algebra-opn} and Remark~\ref{rem:algebra-opn}) 
  and: 
$$ \ulq'_{\it pl} = \opn{Join}(\ulm'_{\it pl},\ulp'_{\it pl}) :
P'_{\it pl} \To G_0 \;\mbox{ with }\;
  \tbl(\ulq'_{\it pl}) = \begin{array}{|l|l|}
  \hline
  {\tt ?a1} & {\tt ?a2} \\
  \hline
  {\tt auth1} & {\tt auth2} \\
  {\tt auth2} & {\tt auth1} \\
  \hline
  \end{array} 
  $$
\item
  Assume that there is some operation $\opn{concat}$ that builds
  a string from any given date and string. 
  For each match $m$ in $\ulm_{\it ps}$ let
  $\ulp_m=\{p_m\}:\{?dm\}\To H_{\it ps}$ 
  where $p_m(?dm)=\opn{concat}(m(?d),m(?m))$ 
  and $H_{\it ps}$ is any graph that contains all strings
  $\opn{concat}(?d,?m)$ as nodes.
  Then: 
  $$ \begin{array}{l}
  \ulq_{\it ps} =
  \opn{Merge}(\ulm_{\it ps},(\ulp_m)_{m\in\ulm_{\it ps}}) : 
  P_{\it ps} \cup \{?dm\} \To G_0 \cup H_{\it ps}
  \\ \;\mbox{ with }\;
  \tbl(\ulq_{\it ps}) = \begin{array}{|l|l|l|l|}
  \hline
  {\tt ?a} & {\tt ?m} & {\tt ?d} & {\tt ?dm} \\
  \hline
  {\tt auth1} & {\tt mes1} & {\tt date1} & {\tt date1mes1} \\
  {\tt auth1} & {\tt mes2} & {\tt date2} & {\tt date2mes2} \\
  {\tt auth2} & {\tt mes3} & {\tt date1} & {\tt date1mes3} \\
  {\tt auth3} & {\tt mes4} & {\tt date4} & {\tt date4mes4} \\
  {\tt auth3} & {\tt mes5} & {\tt date4} & {\tt date4mes5} \\
  \hline
  \end{array} 
  \end{array} 
  $$
    \item
  For each match $m$ in $\ulm_{\it ps}$ let
  $\ulp_m=\{p_m\}:\{?r\}\To H_{\it var}$ 
  where $p_m(?r)$ is some fresh variable $?r_m$
  and $H_{\it var}$ is any graph that contains all variables $?r_m$ as nodes.
  Then: 
  $$ \begin{array}{l}
  \ulq_{\it var} =
  \opn{Merge}(\ulm_{\it ps},(\ulp_m)_{m\in\ulm_{\it ps}}) : 
  P_{\it ps} \cup \{?r\} \To G_0 \cup H_{\it var}
  \\ \;\mbox{ with }\;
  \tbl(\ulq_{\it var}) = \begin{array}{|l|l|l|l|}
  \hline
  {\tt ?a} & {\tt ?m} & {\tt ?d} & {\tt ?r} \\
  \hline
  {\tt auth1} & {\tt mes1} & {\tt date1} & {\tt ?r1} \\
  {\tt auth1} & {\tt mes2} & {\tt date2} & {\tt ?r2} \\
  {\tt auth2} & {\tt mes3} & {\tt date1} & {\tt ?r3} \\
  {\tt auth3} & {\tt mes4} & {\tt date4} & {\tt ?r4} \\
  {\tt auth3} & {\tt mes5} & {\tt date4} & {\tt ?r5} \\
  \hline
  \end{array} 
  \end{array} 
  $$
 \end{itemize}
\end{example}

\begin{definition}[Restrict]
\label{def:algebra-restrict}
Let $\ulm:X\To G$ be a set of matches. 
For every graph $Y$ contained in $X$ and every graph $H$ contained in $G$
such that $\ulm(Y)\subseteq H$, the \emph{restriction}
$\opn{Restrict}(\ulm,Y,H):Y\To H$ 
is made of the restrictions of the matches in $\ulm$
as matches from $Y$ to $H$.
When $H=G$ the notation may be simplified:
$\opn{Restrict}(\ulm,Y)=\opn{Restrict}(\ulm,Y,G):Y\To G$.  
\end{definition}

\begin{definition}[Extend]
\label{def:algebra-extend}
Let $\ulm:X\To G$ be a set of matches. 
For every graph $H$ containing $G$, the \emph{extension}
$\opn{Extend}(\ulm,H) : X\To H$
is made of the extensions of the matches in $\ulm$
as matches from $X$ to $H$.
\end{definition}

New operations are obtained by combining the previous ones
(assuming that $\true$ is a constant).
Comments on Definition~\ref{def:algebra-opn} are given in
Remark~\ref{rem:algebra-opn}. 
We will see in Section~\ref{ssec:gral-pattern} that these derived operations
provide the semantics of the operators of the language $\gral$.

\begin{definition}[derived operations]\
\label{def:algebra-opn}

\begin{itemize}

\item 
  For every sets of matches $\ulm:X\To G$ and $\ulp:Y\To H$,
  let $\ulp_m=\ulp$ for each $m\in\ulm$, then: 
\\ \hsp $\opn{Join}(\ulm,\ulp)
= \opn{Merge}(\ulm,(\ulp_m)_{m\in\ulm})
: X \cup Y \To G\cup H$.

\item For every set of matches $\ulm:X\To G$,
  every family of constants $\ulc=(c_m)_{m\in\ulm}$ and every variable $x$,
  let $\ulp_m=\{p_m\}$ and $p_m(x)=c_m$ for each $m\in\ulm$, then: 
\\ \hsp $\opn{Bind}(\ulm,\ulc,x)
= \opn{Merge}(\ulm,(\ulp_m)_{m\in\ulm}) 
: X \cup \{x\} \To G\cup\ulc$.

\item For every set of matches $\ulm:X\To G$ and 
every family of constants $\ulc=(c_m)_{m\in\ulm}$, 
for some fresh variable $x$,
let $\underline{\true}=(\true)_{m\in\ulm}$: 
\\ \hsp $\opn{Filter}(\ulm,\ulc) 
= \opn{Restrict} (\opn{Bind} (\opn{Bind}(\ulm,\ulc,x),\underline{\true},x),X,G)
: X \To G $.

\item For every set of matches $\ulm:X\To G$ and every graph $R$,
for every $m\in\ulm$ let $p_m:R\to p_m(R)$ be the match such that:
\\ \hsp\hsp $p_m(x)=m(x)$ if $x\in\V(R)\cap\V(X)$
\\ \hsp\hsp and $p_m(x)=\var{x}{m}$ is a fresh variable if $x\in\V(R)\setminus\V(X)$
\\ and let $\ulp_m=\{p_m\}$ and $\ulp(R)=\cup_{m\in\ulm} (p_m(R))$,
then: 
\\ \hsp $\opn{Construct}(\ulm,R)
= \opn{Restrict}( \opn{Merge}(\ulm,(\ulp_m)_{m\in\ulm}),R)
: R \To G\cup \ulp(R) $. 

\item For every sets of matches $\ulm:X\To G$ and $\ulp:X\To H$:
\\ \hsp $\opn{Union}(\ulm,\ulp)
= \opn{Extend}(\ulm,G\cup H) \;\cup\; \opn{Extend}(\ulp,G\cup H)
: X \To G\cup H $.
    
\end{itemize}
\end{definition}

\begin{remark}
\label{rem:algebra-opn}
Let us analyse these definitions.
Note that the definition of $\opn{Bind}$ and $\opn{Filter}$
rely on the fact that isolated nodes are allowed in graphs.

\begin{itemize}

\item 
Operation $\opn{Join}$ is $\opn{Merge}$ when 
the set of matches $\ulp_m$ does not depend on $m$, so that:
\\ \hsp $\opn{Join}(\ulm,\ulp) 
=\{ m\bowtie p \mid m\in\ulm \lwedge p\in\ulp \lwedge m\sim p\}
: X\cup Y \To G\cup H$.
\\ It follows that $\opn{Join}$ is commutative.

\item 
Operation $\opn{Bind}$ is $\opn{Merge}$ when
$\ulp_m$ has exactly one element $p_m$ for each $m$,
which is such that $p_m(x)=c_m$. There are two cases:
\begin{itemize}
\item If $x\in\V(X)$ then this operation selects the matches $m$ in $\ulm$
such that $m(x)=c_m$: 
\\ \hsp $\opn{Bind}(\ulm,\ulc,x) 
=\{ m \mid m\in\ulm \lwedge m(x)=c_m\}
: X \To G$. 
\item If $x\not\in\V(X)$ then this operation extends 
each match $m$ in $\ulm$ by assigning the value $c_m$ to the variable $x$.
Let us denote the resulting match as $m\uplus(x\mapsto c_m)$, so that:
\\ \hsp $\opn{Bind}(\ulm,\ulc,x) 
=\{ m\uplus(x\mapsto c_m) \mid m\in\ulm \}
: X\cup\{x\} \To G\cup \{\ulc\}$. 
\end{itemize}

\item 
Operation $\opn{Filter}$ applies $\opn{Bind}$ twice,
first when $x\not\in\V(X)$ for extending each $m\in\ulm$
by assigning $c_m$ to $x$, then since $x\in\V(X\cup\{x\})$ for selecting
the matches $m$ in $\ulm$ such that $c_m=\true$.
Now the value of the auxiliary variable $x$ is always $\true$, so that $x$
can be dropped: this is the role of the last step which restricts
the domain of the matches from $X\cup\{x\}$ to $X$
and its range from $G\cup\{\ulc\}$ to $G$.

\item 
The first step in operation $\opn{Construct}$ is $\opn{Merge}$ when
$\ulp_m$ has exactly one element $p_m$ for each $m$ (as for $\opn{Bind}$), 
which is determined by $p_m(x)=\var{x}{m}$ for each variable $x$ in $R$
that does not occur in $X$.
Each $\var{x}{m}$ is a fresh variable, which means that 
it is distinct from the variables in $G$, $X$ and $R$,
and that the variables $\var{x}{m}$ are pairwise distinct.
Note that the precise symbol used for denoting $\var{x}{m}$ does not matter.
The second step in operation $\opn{Construct}$ restricts
the domain of the matches from $X\cup R$ to $R$. Thus:
\\ \hsp $\opn{Construct}(\ulm,R) $
is the set of matches from $R$ to $G\cup \ulp(R)$ 
\\ \hsp\hsp determined by the functions $f_m:\V(R)\to\Ll$ 
(for each $m\in\ulm$) such that
\\ \hsp\hsp $f_m(x)=m(x)$ if $x\in\V(R)\cap\V(X)$ and
$f_m(x)=\var{x}{m}$ if $x\in\V(R)\setminus\V(X)$.
\\ Thus, the graph $G\cup \ulp(R)$ is obtained by ``gluing'' one copy of
$G$ with $\Card{\ulm}$ copies of $R$ in the right way. 
Note that the functions $f_m$ are pairwise distinct when $\V(R)$ is not
included in $\V(X)$, but it needs not be the case in general. 
Also, note that the domain $R$ of $\opn{Construct}(\ulm,R)$ may be quite
different from the domain $X$ of $\ulm$, whereas every other operation in
Definition~\ref{def:algebra-opn} either keeps or extends the domain
of $\ulm$. 

\item Operation $\opn{Union}$ is simply the set-theoretic union of
sets of matches which share the same domain (by assumption)
and the same range (by extending the range if necessary). 
This operation differs from the previous ones in the sense that it is not
defined by examining the matches in its arguments.
Note that $\opn{Union}$ is commutative.

\end{itemize}
\end{remark}

\begin{proposition}
\label{prop:algebra-operation}
The sets of matches obtained by the operations previously defined
in this Section have bounded cardinals, as follows. 
\\ \hsp $ \begin{array}{l}
  \Card{\opn{Merge}(\ulm,(\ulp_m)_{m\in\ulm})} \leq \sum_{m\in\ulm}(\Card{\ulp_m})\\
  \Card{\opn{Restrict}(\ulm,X,G)} \leq \Card{\ulm} \\
  \Card{\opn{Extend}(\ulm,H)} = \Card{\ulm} \\
  \Card{\opn{Join}(\ulm,\ulp)} \leq \Card{\ulm}\times\Card{\ulp} \\
  \Card{\opn{Bind}(\ulm,\ulc,x)} = \Card{\ulm} \\
  \Card{\opn{Filter}(\ulm,\ulc)} \leq \Card{\ulm} \\
  \Card{\opn{Construct}(\ulm,R)} \leq \Card{\ulm} \\ 
  \Card{\opn{Union}(\ulm,\ulp)} \leq \Card{\ulm}+\Card{\ulp} \\
\end{array} $
\end{proposition}

The proof of Proposition~\ref{prop:algebra-operation} follows easily
from the definitions.

\section{The Graph Algebraic Query Language} 
\label{sec:gral}

In this Section we introduce the syntax and semantics
of the Graph Algebraic Query Language $\gral$.
There are three syntactic categories in $\gral$:
\emph{expressions}, \emph{patterns} and \emph{queries}.
Expressions are considered in Section~\ref{ssec:gral-expr}.
Patterns are defined in Section~\ref{ssec:gral-pattern},
their semantics is presented as an evaluation function
which maps every pattern $P$ and graph $G$ to a set of matches $\sem{P}{G}$.
Queries are defined in Section~\ref{ssec:gral-query}, 
they are essentially specific kinds of patterns and their semantics
is easily derived from the semantics of patterns, 
the main difference is that the execution of a query on a graph 
returns simply a graph instead of a set of matches.

To each expression $e$ or pattern $P$ is associated a set of variables 
called its \emph{in-scope variables} and denoted $\V(e)$ or $\V(P)$,
respectively. 
An expression $\expr$ is \emph{over} a pattern $P$ if
$\V(\expr)\subseteq\V(P)$. 
In this Section, as in Section~\ref{sec:algebra},  
the set of \emph{labels} $\Ll$ is the union of the disjoint sets
$\mC$ and $\V$, of \emph{constants} and \emph{variables} respectively.
We assume that the set $\mC$ of constants contains the numbers
and strings and the boolean values $\true$ and $\false$,
as well as a symbol $\err$ for errors. 

\subsection{Expressions} 
\label{ssec:gral-expr}
The expressions of $\gral$ are built from the labels using operators, 
which are classified as either basic operators (unary or binary)
and aggregation operators (always unary). 
Remember that typing constraints are not considered in this paper.
Typically, and not exclusively, the sets $\Op_1$, $\Op_2$ and $\Agg$
of \emph{basic unary} operators, \emph{basic binary} operators 
and \emph{aggregation} operators can be: 
\\ \hsp $ \Op_1=\{-,\mathrm{NOT}\}\,,$
\\ \hsp $ \Op_2=\{+,-,\times,/,=,>,<,\mathrm{AND},\mathrm{OR}\}\,,$
\\ \hsp $\Agg = \Agg_{\elem} \cup \{\agg\;\mathrm{DISTINCT} \mid \agg\in\Agg_{\elem}\} \,.$
\\ \hsp\hsp where $\Agg_{\elem} =
\{\mathrm{MAX},\mathrm{MIN},\mathrm{SUM},\mathrm{AVG},\mathrm{COUNT}\}$
\\ A \emph{group of expressions} is a non-empty finite list of expressions. 

\begin{definition}[syntax of expressions]\ 
\label{def:gral-syn-expr} 
The \emph{expressions} $\expr$ of $\gral$ and their set of \emph{in-scope}
variables $\V(\expr)$ are defined recursively as follows:   
\begin{itemize}
\item A constant $c\in\mC$ is an expression 
  with $\V(c)=\emptyset$. 
\item A variable $x\in\V$ is an expression 
  with $\V(x)=\{x\}$. 
\item If $\expr_1$ is an expression and $\op\in\Op_1$
  then $\op\;\expr_1$ is an expression
  with $\V(\op\;\expr_1)=\V(\expr_1)$. 
\item If $\expr_1$ and $\expr_2$ are expressions and $\op\in\Op_2$ 
  then $\expr_1\;\op\;\expr_2$ is an expression
  with $\V(\expr_1\;\op\;\expr_2)=\V(\expr_1)\cup\V(\expr_2)$.
\item If $\expr_1$ is an expression and $\agg\in\Agg$
  then $\agg(\expr_1)$ is an expression
  with $\V(\agg(\expr_1))=\V(\expr_1)$. 
\item If $\expr_1$ is an expression, $\agg\in\Agg$ and $\gp$
  a group of expressions with all its variables distinct from the
  variables in $\expr_1$, 
  then $\agg(\expr_1\mbox{ {\rm BY} }\gp)$ is an expression
  with $\V(\agg(\expr_1\mbox{ {\rm BY} }\gp))=\V(\expr_1)$.
\end{itemize}
\end{definition}

The \emph{value} of an expression with respect to a set of matches $\ulm$ 
(Definition~\ref{def:gral-sem-expr}) is a family of constants 
$\ev(\ulm,\expr)=(\ev(\ulm,\expr)_m)_{m\in\ulm}$ indexed by the set $\ulm$.
Each constant $\ev(\ulm,\expr)_m$ depends on $\expr$ and $m$ 
and it may also depend on other matches in $\ulm$
when $\expr$ involves aggregation operators.
The \emph{value} of a group of expressions $\gp=(\expr_1,...,\expr_k)$ 
with respect to $\ulm$ is the list $\ev(\ulm,\gp)_m=
(\ev(\ulm,\expr_1),...,\ev(\ulm,\expr_k))$. 
To each basic operator $\op$ is associated a function $\deno{op}$
(or simply $\op$) 
from constants to constants if $\op$ is unary
and from pairs of constants to constants if $\op$ is binary.
To each aggregation operator $\agg$ in $\Agg$ is associated a function
$\deno{agg}$ (or simply $\agg$)
from \emph{multisets} of constants to constants. 
Note that each family of constants determines a multiset of constants: 
for instance a family $\ulc=(c_m)_{m\in\ulm}$ of constants indexed by the
elements of a set of matches $\ulm$ determines the multiset of constants
$\bag{c_m\mid m\in\ulm}$, 
which is also denoted $\ulc$ when there is no ambiguity.
Some aggregation operators $\agg$ in $\Agg_{\elem}$ are such that 
$\deno{agg}(\ulc)$ depends only on the set underlying the multiset $\ulc$, 
which means that $\deno{agg}(\ulc)$ 
does not depend on the multiplicities in the multiset $\ulc$:
for instance this is the case for MAX and MIN 
but not for SUM, AVG, COUNT.
When $\agg=\agg_{\elem}\;\mathrm{DISTINCT}$ with $\agg_{\elem}$ in $\Agg_{\elem}$
then $\deno{agg}(\ulc)$ is $\deno{agg_{\elem}}$ applied to
the underlying set of $\ulc$.
For instance, $\clause{COUNT }(\ulc)$ counts the number of elements
of the multiset $\ulc$ with their multiplicies, while 
$\clause{COUNT DISTINCT }(\ulc)$ counts the number of distinct elements
in $\ulc$. 

\begin{definition}[evaluation of expressions]
\label{def:gral-sem-expr}
Let $X$ be a graph, $\expr$ an expression over $X$ and $\ulm:X\To Y$
a set of matches. 
The \emph{value} of $\expr$ with respect to $\ulm$ is the family of constants
$\ev(\ulm,\expr) = (\ev(\ulm,\expr)_m)_{m\in\ulm}$ defined recursively
as follows (with notations as in Definition~\ref{def:gral-syn-expr}):
\begin{itemize}
  \item $\ev(\ulm,c)_m = c$. 
  \item $\ev(\ulm,x)_m = m(x)$.
\item $\ev(\ulm,\op\;\expr_1)_m = \deno{\op}\,\ev(\ulm,\expr_1)_m\,$.  
\item $\ev(\ulm,\expr_1\;\op\;\expr_2)_m =
  \ev(\ulm,\expr_1)_m\,\deno{\op}\,\ev(\ulm,\expr_2)_m\,$. 
\item $\ev(\ulm,\agg(\expr_1))_m = \deno{\agg}(\ev(\ulm,\expr_1))$
  (which is the same for every $m$ in $\ulm$). 
\item $\ev(\ulm,\agg(\expr_1\;BY\;\gp))_m =
  \deno{\agg}(\ev(\ulm|_{\gp,m},\expr_1))$ 
  where $\ulm|_{\gp,m}$ is the subset of $\ulm$ 
  made of the \\ matches $m'$ in $\ulm$ such that
  $ \ev(\ulm,\gp)_{m'} = \ev(\ulm,\gp)_m$ 
  (which is the same for every $m$ and $m'$ in \\ $\ulm$
  such that $\ev(\ulm,\gp)_m=\ev(\ulm,\gp)_{m'}$). 
\end{itemize} 
\end{definition}

\begin{example}
  \label{ex:gral-sem-expr}
  Let $\ulm_{\it pl}$ be as in Example~\ref{ex:set-of-matches}.
  For every $m$ in $\ulm_{\it pl}$ we have: 
  $$ \ev(\ulm_{\it pl},\clause{COUNT }(likes))_m=5 $$
  whereas $\ev(\ulm_{\it pl},\clause{COUNT }(likes \clause{ BY }?a1))_m$
  depends on the match $m$ in $\ulm_{\it pl}$:
  $$ \ev(\ulm_{\it pl},\clause{COUNT }(likes \clause{ BY }?a1))_m=
  \begin{cases}
    1 & \mbox{ when } m(?a1)={\tt auth1} \\ 
    1 & \mbox{ when } m(?a1)={\tt auth2} \\ 
    3 & \mbox{ when } m(?a1)={\tt auth3} \\ 
    \end{cases}
  $$
\end{example}

\begin{definition}[equivalence of expressions]
\label{def:gral-equiv-expr}
Two expressions over a graph $X$ are \emph{equivalent} if they have
the same value with respect to every set of matches $\ulm:X\To Y$.
\end{definition}

\subsection{Patterns}
\label{ssec:gral-pattern}

In Definition~\ref{def:gral-syn-pattern} the patterns of $\gral$ are built
from graphs by using five operators: 
$\clause{JOIN}$, $\clause{BIND}$, $\clause{FILTER}$,
$\clause{CONSTRUCT}$ and $\clause{UNION}$.
In Definition~\ref{def:gral-sem-pattern} the semantics of patterns
is given by an evaluation function.
  Some patterns have an associated graph called a \emph{template},
  such a pattern $P$ may give rise to a query $Q$
  as explained in Section~\ref{ssec:gral-query}, 
  then the result of query $Q$ is built from the template of $P$. 

\begin{definition}[syntax of patterns]
  \label{def:gral-syn-pattern}
The \emph{patterns} $P$ of $\gral$, their set of \emph{in-scope}
variables $\V(P)$ and their \emph{template} graph $\T(P)$ when it exists  
are defined recursively as follows.
\begin{itemize}
\item A graph is a pattern, called a \emph{basic pattern},
  and $\V(P)$ is the set of variables of the graph $P$.
\item If $P_1$ and $P_2$ are patterns then
  $P_1\clause{ JOIN }P_2$ is a pattern
  and $\V(P_1\clause{ JOIN }P_2)=\V(P_1)\cup\V(P_2)$.
  \item If $P_1$ is a pattern, $\expr$ an expression over $P_1$ 
    and $x$ a variable 
    then  $P_1\clause{ BIND }\expr\clause{ AS }x$ is a pattern
  and $\V(P_1\clause{ BIND }\expr\clause{ AS }x)=\V(P_1)\cup\{x\}$. 
  \item If $P_1$ is a pattern and $\expr$ an expression over $P_1$ 
  then $P_1\clause{ FILTER }\expr$ is a pattern 
  and $\V(P_1\clause{ FILTER }\expr)=\V(P_1)$.
\item If $P_1$ is a pattern and $R$ a graph 
  then $P_1\clause{ CONSTRUCT }R$, 
  \\ also written $\clause{ CONSTRUCT }R\clause{ WHERE }P_1$, 
  is a pattern  
  and $\V(P_1\clause{ CONSTRUCT }R)=\V(R)$.
  \\ In addition this pattern has a template $\T(P_1\clause{ CONSTRUCT }R)=R$. 
\item If $P_1$ and $P_2$ are patterns with template and if
  $\T(P_1)=\T(P_2) = R$
  then \\ $P_1\clause{ UNION }P_2$ is a pattern
  and $\V(P_1\clause{ UNION }P_2)=\V(R)$.
  \\ In addition this pattern has a template
  $\T(P_1\clause{ UNION }P_2)=\T(P_1)=\T(P_2)$. 
\end{itemize}
\end{definition}

The \emph{value} of a pattern over a graph is a set of matches, as defined now.

\begin{definition}[evaluation of patterns]
\label{def:gral-sem-pattern}
The \emph{value} of a pattern $P$ of $\gral$ over 
a graph $G$ is a set of matches $\sem{P}{G}:\se{P}\To\seg{P}{G}$
from a graph $\se{P}$ that depends only on $P$ 
to a graph $\seg{P}{G}$ that contains $G$. 
This value $\sem{P}{G}:\se{P}\To\seg{P}{G}$ is 
defined inductively as follows
(with notations as in Definition~\ref{def:gral-syn-expr}):
\begin{itemize}
\item If $P$ is a basic pattern then 
  $\sem{P}{G}
  = \Ulm(P,G)
  :P\To G $.
\item $\sem{P_1\clause{ JOIN }P_2}{G}
  = \opn{Join}(\sem{P_1}{G},\sem{P_2}{\seg{P_1}{G}})
  : \se{P_1}\cup\se{P_2}\To \seg{P_2}{\seg{P_1}{G}}$.
\item $\sem{P_1\clause{ BIND }\expr\clause{ AS }x}{G}
  = \opn{Bind}(\sem{P_1}{G},\ev(\sem{P_1}{G},\expr),x)
  : \se{P_1}\cup\{x\}\To \seg{P_1}{G}\cup{\sem{P_1}{G}(\expr)} $.
\item $\sem{P_1\clause{ FILTER }\expr}{G}
  = \opn{Filter}(\sem{P_1}{G},\ev(\sem{P_1}{G},\expr))
  : \se{P_1}\To \seg{P_1}{G} $.
\item $\sem{P_1\clause{ CONSTRUCT }R}{G}
  = \opn{Construct}(\sem{P_1}{G},R)
  : R\To \seg{P_1}{G} \cup \sem{P_1}{G}(R)$. 
\item $\sem{P_1\clause{ UNION }P_2}{G} 
  = \opn{Union}(\sem{P_1}{G},\sem{P_2}{\seg{P_1}{G}})
  : R \To \seg{P_2}{\seg{P_1}{G}} $
  where $R=\T(P_1)=\T(P_2)$.
\end{itemize}
\end{definition}

\begin{remark} 
\label{rem:gral-sem-pattern}
Note that, syntactically, each operator $\clause{OP}$ builds a pattern $P$
from a pattern $P_1$ and a parameter $\param$,
which is either a pattern $P_2$ (for $\clause{JOIN}$ and $\clause{UNION}$),
a pair $(\expr,x)$ made of an expression and a variable (for $\clause{BIND}$),
an expression $\expr$ (for $\clause{FILTER}$)
or a graph $R$ (for $\clause{CONSTRUCT}$).
Semantically, for every non-basic pattern $P=P_1\clause{ OP }\param$,
we denote $\ulm_1:X_1\To G_1$ for $\sem{P_1}{G}:\se{P_1}\To\seg{P_1}{G}$
and $\ulm:X\To G'$ 
for $\sem{P}{G}:\se{P}\To\seg{P}{G}$. 
In every case it is necessay to evaluate $\ulm_1$ before evaluating $\ulm$:
for $\clause{JOIN}$ and $\clause{UNION}$ this is because pattern $P_2$
is evaluated on $G_1$,
for $\clause{BIND}$ and $\clause{FILTER}$ because expression $\expr$
is evaluated with respect to $\ulm_1$,
and for $\clause{CONSTRUCT}$ because of the definition of $\opn{Construct}$.
According to Definition~\ref{def:gral-sem-pattern} 
given a pattern $P$ and a graph $G$, 
the value $\ulm:X\To G'$ of $P$ is determined as follows:
\begin{itemize}
\item When $P$ is a basic pattern then $X=P$, $G'=G$ and 
  $\ulm$ is made of all matches from $P$ to $G$.
\item $P=P_1\clause{ OP }\param$ then the semantics of $P$
  is easily derived from Definition~\ref{def:algebra-opn}
  (see also Remark~\ref{rem:algebra-opn}). However, note that the
  semantics of $P_1\clause{ JOIN }P_2$ and $P_1\clause{ UNION }P_2$ is
  not symmetric in $P_1$ and $P_2$ in general, unless $\seg{P_1}{G}=G$
  and $\seg{P_2}{G}=G$, which occurs when $P_1$ and $P_2$ are basic patterns.
\item The graph $\seg{P}{G}$ is built by adding to $G$
  ``whatever is required'' for the evaluation,
  in examples we often avoid its precise description.  
\end{itemize}
Given a non-basic pattern $P=P_1\clause{ OP }\param$, the pattern 
$P_1$ is a \emph{subpattern} of $P$, as well as $P_2$ when
$P=P_1\clause{ JOIN }P_2$ or $P=P_1\clause{ UNION }P_2$.
The semantics of patterns is defined in terms of the semantics of
its subpatterns (and the semantics of its other arguments, if any).
Thus, for instance, CONSTRUCT patterns can be nested at any depth.
\end{remark}

\begin{proposition} 
\label{prop:gral-sem-pattern}
For every pattern $P$, the set $\V(P)$ of in-scope variables of $P$
is the same as the set $\V(\se{P})$ of variables of the graph $\se{P}$.
\end{proposition}

\begin{example}
\label{ex:gral-sem-pattern}

In each item below we consider first some pattern $P_i$
and some template $R_i$, then the pattern
$$ C_i=P_i\clause{ CONSTRUCT }R_i \;\mbox{ also written as }\; 
C_i=\clause{CONSTRUCT }R_i\clause{ WHERE }P_i \;.$$
We refer to Examples~\ref{ex:set-of-matches} and~\ref{ex:gral-sem-expr}.

\begin{itemize}
  
  \item
\begin{verbatim}
  C1 = CONSTRUCT { ?a1 cites ?a2 }
       WHERE { ?a1 publishes ?m1 . ?m1 refersTo ?m2 . ?a2 publishes ?m2 }
\end{verbatim}
Here, $C_1=P_1\clause{ CONSTRUCT }R_1$ where $P_1=P_{\it prp}$, 
so that the value of $P_1$ over $G_0$ is $\ulm_{\it prp}:P_1\To G$.
Note that $\V(R_1)\subseteq\V(P_1)$. 
Let $G_1= G_0 \cup
\{  {\tt auth2} \;{\tt cites}\; {\tt auth1} \;.\;
    {\tt auth3} \;{\tt cites}\; {\tt auth1} \}$,
the value of $C_1$ over $G_0$ is: 
$$ \sem{C_1}{G_0}:R_1\To G_1 \;\mbox{ with }\; 
  \tbl(\sem{C_1}{G_0}) = \begin{array}{|l|l|}
  \hline
  {\tt ?a1} & {\tt ?a2} \\
  \hline
  {\tt auth2} & {\tt auth1} \\
  {\tt auth3} & {\tt auth1} \\
  \hline
  \end{array} 
$$

  \item
\begin{verbatim}
  C2 = CONSTRUCT { ?n }
       WHERE {
         ?a likes ?m 
         BIND COUNT(likes) AS ?n }
\end{verbatim}
Here, $C_2=P_2\clause{ CONSTRUCT }R_2$ where 
the template $R_2$ is the graph made of only one isolated node 
which is the variable $?n$. 
The graph $\se{P_2}$ is $\{ {\tt ?a} \;{\tt likes}\; {\tt ?m} \;.\;{\tt ?n} \}$.
Let $G_2= G_0 \cup\{5\}$, 
the value of $C_2$ over $G_0$ is: 
$$ \sem{P_2}{G_0} : \se{P_2} \To G_2 \;\mbox{ with }\;
\tbl(\sem{P_2}{G_0}) = 
  \begin{array}{|l|l|l|}
  \hline
  {\tt ?a} & {\tt ?m} & {\tt ?n} \\
  \hline
  {\tt auth1} & {\tt mes3} & {\tt 5} \\
  {\tt auth1} & {\tt mes4} & {\tt 5} \\
  {\tt auth1} & {\tt mes5} & {\tt 5} \\
  {\tt auth2} & {\tt mes1} & {\tt 5} \\
  {\tt auth2} & {\tt mes4} & {\tt 5} \\
  \hline
  \end{array} 
$$
  Then the graph $\se{C_2}$ is $\{ {\tt ?n} \}$
  and the value of $C_2$ over $G_0$ is: 
$$ \sem{C_2}{G_0} : \se{C_2} \To G_2 \;\mbox{ with }\;
  \tbl(\sem{C_2}{G_0}) = \begin{array}{|l|}
  \hline
  {\tt ?n} \\
  \hline
  {\tt 5} \\
  \hline
  \end{array} 
$$

  \item
\begin{verbatim}
  C3 = CONSTRUCT { ?a1 nbOfLikes ?n }
       WHERE { ?a1 publishes ?m . ?a2 likes ?m 
         FILTER (NOT(?a1=?a2)) 
         BIND COUNT(likes BY ?a1) AS ?n } 	
\end{verbatim}
Here, $C_3=P_3\clause{ CONSTRUCT }R_3$ where 
$R_3=\{ {\tt ?a1}  \; {\tt nbOfLikes} \; {\tt ?n} \}$ is made of
one triple 
and $\se{P_3} = \{ {\tt ?a1} \;{\tt publishes}\; {\tt ?m} \;.\;
{\tt ?a2} \;{\tt likes}\; {\tt ?m} \;.\; {\tt ?n} \}$.
The evaluation of $C_3$ over $G_0$ starts from $\ulm_{\it pl}$. 
Let $G_3= G_0 \cup \{ {\tt auth1}\;{\tt nbOfLikes}\;{\tt 1}
\;.\; {\tt auth2}\;{\tt nbOfLikes}\;{\tt 1} 
\;.\; {\tt auth3}\;{\tt nbOfLikes}\;{\tt 3} \}$, 
the value of $C_3$ over $G_0$ is: 
$$ \sem{P_3}{G_0} : \se{P_3} \To G_3 \;\mbox{ with }\;
\tbl(\sem{P_3}{G_0}) = 
  \begin{array}{|l|l|l|l|}
  \hline
  {\tt ?a1} & {\tt ?m} & {\tt ?a2} & {\tt ?n} \\
  \hline
   {\tt auth1} & {\tt mes1} & {\tt auth2} & {\tt 1} \\
   {\tt auth2} & {\tt mes3} & {\tt auth1} & {\tt 1} \\
   {\tt auth3} & {\tt mes4} & {\tt auth1} & {\tt 3} \\
   {\tt auth3} & {\tt mes4} & {\tt auth2} & {\tt 3} \\
   {\tt auth3} & {\tt mes5} & {\tt auth1} & {\tt 3} \\
  \hline
  \end{array} 
$$
  Then the graph $\se{C_3}$ is
  $R_3=\{ {\tt ?a1}  \; {\tt nbOfLikes}\; {\tt ?n} \}$
  and the value of $C_3$ over $G_0$ is: 
$$ \sem{C_3}{G_0} : \se{C_3} \To G_3 \;\mbox{ with }\;
  \tbl(\sem{C_3}{G_0}) = \begin{array}{|l|l|}
  \hline
  {\tt ?a1} & {\tt ?n} \\
  \hline
   {\tt auth1} & {\tt 1} \\
   {\tt auth2} & {\tt 1} \\
   {\tt auth3} & {\tt 3} \\
  \hline
  \end{array} 
  $$

\item
  \begin{verbatim}
  C4 = CONSTRUCT { ?a1 nbOfFriends ?n }
       WHERE 
       { 
         CONSTRUCT { ?a1 friend ?a2 } 
         WHERE {
           ?a1 publishes ?m1 . ?a2 likes ?m1 .
           ?a2 publishes ?m2 . ?a1 likes ?m2   
         }
         BIND COUNT (friend BY ?a1) AS ?n     
       }
\end{verbatim}
  Here $C_4=P_4\clause{ CONSTRUCT }R_4$ where $P_4$ itself
  contains a subpattern $C'_4=P'_4\clause{ CONSTRUCT }R'_4$.
The evaluation of the basic pattern $P'_4$ over $G_0$ gives
$$ \sem{P'_4}{G_0} : P'_4 \To G_0  \;\mbox{ with }\;
  \tbl(\sem{P'_4}{G_0}) = \begin{array}{|l|l|l|l|}
  \hline
  {\tt ?a1} & {\tt ?m1} & {\tt ?a2} & {\tt ?m2} \\
  \hline
   {\tt auth1} & {\tt mes1} & {\tt auth2} & {\tt mes3} \\
   {\tt auth2} & {\tt mes3} & {\tt auth1} & {\tt mes1} \\
  \hline
  \end{array} 
$$
Then we get $\se{C'_4} = \{ {\tt ?a1} \;{\tt friend}\; {\tt ?a2} \}$
and: 
$$ \sem{C'_4}{G_0} : \se{C'_4} \To \seg{C'_4}{G_0}  \;\mbox{ with }\;
  \tbl(\sem{C'_4}{G_0}) = \begin{array}{|l|l|}
  \hline
  {\tt ?a1} & {\tt ?a2} \\
  \hline
   {\tt auth1} & {\tt auth2} \\
   {\tt auth2} & {\tt auth1} \\
  \hline
  \end{array} 
  $$
Finally $\se{C_4} = \{ {\tt ?a1} \;{\tt NbOfFriends}\; {\tt ?n} \}$
and: 
$$ \sem{C_4}{G_0} : \se{C_4} \To  \seg{C_4}{G_0}  \;\mbox{ with }\;
  \tbl(\sem{C_4}{G_0}) = \begin{array}{|l|l|}
  \hline
  {\tt ?a1} & {\tt ?n} \\
  \hline
   {\tt auth1} & {\tt 1} \\
   {\tt auth2} & {\tt 1} \\
  \hline
  \end{array} 
  $$

\item
\begin{verbatim}
  C5 = CONSTRUCT { ?r author ?a . ?r date ?d }
       WHERE { ?a publishes ?m . ?m stampedAt ?d }  
\end{verbatim}
Here $C_5=P_5\clause{ CONSTRUCT }R_5$ with a variable $?r$ in $R_5$
that does not occur in $P_5$. 
The value of the basic pattern $P_5$ over $G_0$ is:
$$ \sem{P_5}{G_0}: P_5 \To G_0 \;\mbox{ with }\;
  \tbl(\sem{P_5}{G_0}) = \begin{array}{|l|l|l|}
  \hline
  {\tt ?a} & {\tt ?m} & {\tt ?d} \\
  \hline
  {\tt auth1} & {\tt mes1} & {\tt date1} \\
  {\tt auth1} & {\tt mes2} & {\tt date2} \\
  {\tt auth2} & {\tt mes3} & {\tt date1} \\
  {\tt auth3} & {\tt mes4} & {\tt date4} \\
  {\tt auth3} & {\tt mes5} & {\tt date4} \\
  \hline
  \end{array} 
  $$
  Then $\se{C_5} = R_5$ and, since $\sem{P_5}{G_0}$ is made of 5 matches,
  the value $\sem{C_5}{G_0}$ is obtained by gluing 5 copies of $R_5$,
  with 5 fresh variables corresponding to different renamings of
  variable $?r$. 
Indeed,  the variable $?r$ in $R_5$, which is not a variable of $P_5$,
gives rise to one fresh variable for each match of $P_5$ in $G_0$. Thus:
$$ \sem{C_5}{G_0} : R_5 \To \seg{C_5}{G_0}  \;\mbox{ with }\;
  \tbl(\sem{C_5}{G_0}) = \begin{array}{|l|l|l|}
  \hline
  {\tt ?r} & {\tt ?a} & {\tt ?d} \\
  \hline
  {\tt ?r1} & {\tt auth1} & {\tt date1} \\
  {\tt ?r2} & {\tt auth1} & {\tt date2} \\
  {\tt ?r3} & {\tt auth2} & {\tt date1} \\
  {\tt ?r4} & {\tt auth3} & {\tt date4} \\
  {\tt ?r5} & {\tt auth3} & {\tt date4} \\
  \hline
  \end{array} 
$$
   
\end{itemize}
\end{example}

\begin{definition}[equivalence of patterns] 
\label{def:gral-equiv-pattern}
Two patterns are \emph{equivalent} if they have the same value
over $G$ for every graph $G$. 
\end{definition}

\begin{proposition} 
\label{prop:gral-sem-basic}
For every basic patterns $P_1$ and $P_2$, the basic pattern $P_1\cup P_2$
is equivalent to $P_1 \clause{ JOIN }P_2$ and to $P_2 \clause{ JOIN }P_1$.
\end{proposition}

\subsection{Queries}
\label{ssec:gral-query}

A query in $\gral$ is essentially a pattern which has a template.
The main difference between patterns and queries is that,
while a pattern is interpreted as a function from graphs to sets of matches,
a query is interpreted as a function from graphs to graphs.
The operator for building queries from patterns is denoted $\clause{GRAPH}$.
According to Definition~\ref{def:gral-sem-pattern}, the value of
a pattern $P$ with template $R$ over a graph $G$ is a set of matches
$\sem{P}{G}:R\To \seg{P}{G}$, and
the semantics of patterns is defined recursively in terms of their values.
Thus, patterns have a graph-to-set-of-matches semantics,
while queries have a graph-to-graph semantics, as defined below,
based on Definition~\ref{def:algebra-image-s} of
the image of a graph by a set of functions. 

\begin{definition}[syntax of queries]
\label{def:gral-syn-query}
A \emph{query} $Q$ of $\gral$ is written $\clause{GRAPH }(P)$ 
where $P$ is either a $\clause{CONSTRUCT}$ or a $\clause{UNION}$ pattern. 
Then \emph{the pattern of} $Q$ is $P$
and the \emph{template} $\T(Q)$ of $Q$ is the template of $P$.
\end{definition}

\begin{definition}[result of queries]
\label{def:gral-sem-query}
The \emph{result} of a query $Q$ with pattern $P$ and template $R$
over a graph $G$ 
is the subgraph of $\seg{P}{G}$ image of $R$ by $\sem{P}{G}$,
it is denoted $\Result(Q,G)$. 
\end{definition}

Thus, when $Q=\clause{GRAPH }(P_1\clause{ CONSTRUCT }R)$,
the result of $Q$ over $G$ is the graph $\Result(Q,G)=\sem{P_1}{G}(R)$
built by ``gluing'' the graphs $m(R)$ for $m\in\sem{P_1}{G}$, 
where $m(R)$ is a copy of $R$ with each variable $x\in\V(R)\setminus\V(X)$
replaced by a fresh variable $\var{x}{m}$. 
And when $Q=\clause{GRAPH }(P_1\clause{ UNION }P_2)$,
the result of $Q$ over $G$ is the graph $\Result(Q,G)=H_1\cup H_2$ where
$H_i=\Result(\clause{GRAPH }(P_i),G)$ and the fresh variables occuring
in $H_1$ are distinct from the ones in $H_2$. 

\begin{example}
\label{ex:gral-sem-query}

It is now easy to compute the result of the $\gral$ queries: 
$$ Q_i=\clause{GRAPH }(C_i) $$
over $G_0$ when $C_i$ is a pattern from Example~\ref{ex:gral-sem-pattern}. 
We know that the result of $Q_i$ applied to $G_0$
is an instance of $R_i$ when $\V(R_i)\subseteq \V(C_i)$,
and that in general it is built by ``gluing'' together several instances
of $R_i$ (as for query $Q_5$ below). 

\begin{itemize}
  
  \item {\bf Author citations.} 
Let us say that an author $a1$ {\em cites} an author $a2$ 
when $a1$ has published a message that refers to a message published by $a2$.
In order to build the graph of author citations we use the query:
\\ \hsp $Q_1=\clause{GRAPH }(C_1)$.
\\ {F}rom $ \sem{C_1}{G_0}$ we get the graph: 
$$ \Result(Q_1,G_0) = \{ \; \verb+auth2 cites auth1+ \;.\;
  \verb+auth3 cites auth1+ \;\} \;.$$

  \item {\bf Number of likes.}
Let us count the number of \emph{likes} in the database. 
We can get this result by counting the number of triples with predicate
${\tt likes}$, or equivalently the number of predicates ${\tt likes}$.
We use the query:
\\ \hsp $Q_2=\clause{GRAPH }(C_2)$.
\\ {F}rom $ \sem{C_2}{G_0}$ we get the graph: 
$$ \Result(Q_2,G_0) = \{ \; \verb+5+ \;\} $$
This result is the number $5$, which is considered in $\gral$ as
a graph made of only one isolated node.
Note that we would get a query equivalent to $Q_2$
by counting either the number of authors $?a$ who like a message 
(with multiplicity the number of messages liked by $?a$),
or by counting the number of messages $?m$ which are liked by someone
(with multiplicity the number of authors who like $?m$).
This means that 
the line \verb+BIND COUNT(likes) AS ?n + 
could be replaced either by \verb+BIND COUNT(?a) AS ?n +
or by \verb+BIND COUNT(?m) AS ?n +.

  \item {\bf Number of likes per author.}
    Let us now count the number of \emph{likes per author} in the database,
    which means, for each author count the 
number of likes of messages published by this author, except for self-likes. 
We display the result as the graph made of the triples
${\tt \;?a1 \;nbOfLikes\; ?n }$ where $?n$ is the
number of likes of author $?a1$, by using the query:
\\ \hsp $Q_3=\clause{GRAPH }(C_3)$.
\\ {F}rom $ \sem{C_3}{G_0}$ we get the graph: 
$$ \Result(Q_3,G_0) =  \{
\verb+ auth1 nbOfLikes 1+ \;.\;
\verb+ auth2 nbOfLikes 1+ \;.\;
\verb+ auth3 nbOfLikes 3+  \;\} \;. $$

\item {\bf Number of friends per author.}
Let us count the number of friends of each author,
where friendship is the symmetric relation between authors
defined as follows: two authors are \emph{friends} when each one likes
a publication by the other (here self-friends are allowed).
We use the query:
\\ \hsp $Q_4=\clause{GRAPH }(C_4)$.
\\ {F}rom $ \sem{C_4}{G_0}$ we get the graph: 
$$ \Result(Q_4,G_0) = \{ \; \verb+auth1 nbOfFriends 1+ \;.\; 
\verb+auth2 nbOfFriends 1+ \; \} \;.$$
  
\item {\bf Generation of fresh variables.}
Now let us build, for each author $?a$ and each message
$?m$ published by $?a$ and stamped at date $?d$,
a tree with a fresh variable as root and with two branches, 
one named ${\tt author}$ towards $?a$ and
the other one named ${\tt date}$ towards $?d$.
We use the query:
\\ \hsp $Q_5=\clause{GRAPH }(C_5)$.
\\ {F}rom $ \sem{C_5}{G_0}$ we get the graph: 
$$ \Result(Q_5,G_0) = T_1 \cup T_2 \cup T_3 \cup T_4 \cup T_5 $$
where each $T_i$ is the copy of $R_5$ corresponding to the $i$-th row
in $\tbl(\sem{C_5}{G_0})$, so that: 
$$ \begin{array}{l}
T_1 = \{ \verb+ ?r1 author auth1 . ?r1 date date1+ \; \} \\ 
T_2 = \{ \verb+ ?r2 author auth1 . ?r2 date date2+ \; \} \\ 
T_3 = \{ \verb+ ?r3 author auth2 . ?r3 date date1+ \; \} \\ 
T_4 = \{ \verb+ ?r4 author auth3 . ?r4 date date4+ \; \} \\ 
T_5 = \{ \verb+ ?r5 author auth3 . ?r5 date date4+ \; \} \\ 
\end{array} $$

In fact, query $Q_5$ ``mimicks'' the following SELECT query $Q'_5$:
\begin{verbatim}
  SELECT ?a ?d 
  WHERE { ?a publishes ?m . ?m stampedAt ?d }  
\end{verbatim}
As explained in \cite{DEP2020arxiv}, 
the various copies of the variable $?r$ in the result of $Q_5$ 
act as identifiers for the rows in the table result of $Q'_5$
over $G_0$ (as for instance in $\sparql$),
which is obtained by dropping the column $?r$ from $\tbl(\sem{P_5}{G_0})$. 
Note that the table $\tbl(\sem{P_5}{G_0})$ has all its rows distinct by
definition, whereas this becomes false when the column $?r$ is dropped.

\end{itemize}
\end{example}

\begin{definition}[equivalence of queries] 
\label{def:gral-equiv-query} 
Two queries are \emph{equivalent} if they have the same template and
the same result over every graph.
\end{definition}

It follows that queries with equivalent patterns are equivalent, 
but this condition is not necessary. 

\begin{remark}[about SPARQL queries]
\label{rem:gral-sparql}
CONSTRUCT queries in $\sparql$ are similar to CONSTRUCT queries in $\gral$:
the variables in $\V(R)\setminus \V(X)$ in $\gral$ play the same role
as the blank nodes in $\sparql$.
However the subCONSTRUCT patterns are specific to $\gral$.
There is no SELECT query in this core version of $\gral$, however
following \cite{DEP2020arxiv} we may consider SELECT queries as kinds of
CONSTRUCT queries.
\end{remark}

\section{Conclusion}
\label{sec:conclusion}

We considered the problem of the evaluation of graph-to-graph queries,
namely CONSTRUCT queries, possibly involving nested sub-queries. We
proposed a new uniform evaluation semantics of such queries which rests on a
recursive definition of the notion of patterns and a new definition of
the considered graphs which are allowed to have isolated nodes. Hence, 
the evaluation of a pattern always yields a
pair consisting of a graph and a set of matches (variable
assignments).  Notice that we did not tackle explicitly graph-to-table
queries such as the well-known SELECT queries. We have shown recently
in \cite{DEP2020arxiv} that SELECT queries are particular case of
CONSTRUCT queries. This stems from an easy encoding of tables as
graphs. Thus, the proposed semantics can be extended immediately to SELECT
queries involving Sub-SELECT queries.

The present work opens several perspectives including a generalization
of the proposed semantics to other models of graphs such as
\emph{property graphs}. Such an extension needs to ensure the
existence of the main operations of the proposed algebra such as the
$\opn{Merge}$ operation.  An operational semantics, based on rewriting
systems, which is faithful with the evaluation semantics proposed in
this paper is under progress. Its underlying rewrite rules are
inspired by the algebraic approach in \cite{DEP-GCM20}.

Furthermore, the core language $\gral$ contains only simple patterns
needed to illustrate our uniform semantics. Comparison with other
expressions such as EXISTS$(pattern)$ or patterns such as
FROM$(query)$ \cite{AnglesG11,PolleresRK16} remains to be
investigated.


\end{document}